\documentclass[cameraready]{Interspeech}

\usepackage{amsmath, amssymb}
\usepackage{graphicx, xcolor}
\usepackage{multirow, multicol}
\usepackage{url}
\usepackage{adjustbox, colortbl}
\usepackage{booktabs}
\usepackage{subfig}
\usepackage{comment}
\usepackage{hyperref}
\definecolor{lavender}{RGB}{10,100,180}   % soft light 

\definecolor{fern}{RGB}{113,159,117}   % soft fern green
\title{Towards Detecting Neural Audio Codec Synthesized Heart Sounds}

\author[affiliation={1,4}, equalcontribution]{Girish}{}
\author[affiliation={2}, equalcontribution]{Orchid}{Chetia Phukan}
\author[affiliation={3,4}, equalcontribution]{Mohd}{Mujtaba Akhtar}
\author[affiliation={4}, equalcontribution]{Bhavinkumar Vinodbhai}{Kuwar}
\author[affiliation={5}]{Swarup Ranjan}{Behera}
\author[affiliation={4}]{Arun Balaji}{Buduru}
\address{
   $^1$ UPES, India
    $^2$ NTHU, Taiwan
    $^3$ VBSPU, India,
    $^4$ IIIT-Delhi, India 
    $^5$ Independent Researcher, India
}

\email{\textbf{\textcolor{blue}{Correspondence:}} orchidchetiaphukan1@gmail.com
}


\usepackage{comment}
\usepackage{url}
\begin{document}

\maketitle

\begin{abstract}
In this paper, we introduce Synthetic Heart Sound Detection (SHAC), a task aimed at identifying phonocardiograms (PCGs) synthesized using neural audio codecs (NACs). To facilitate research in this direction, we release \textbf{\texttt{CARDIOFAKE}}, the first benchmark dataset for SHAC containing both real and codec-synthesized PCGs. We benchmark spectral representations (MFCC, LFCC) and self-supervised learning (SSL) representations (e.g., WavLM) for the task. Furthermore, we propose \textbf{\texttt{GROOT}}, a fusion framework that integrates spectral and SSL features for leveraging their complementary behavior. Experiments show that \textbf{\texttt{GROOT}}, combining MFCC and WavLM, achieves state-of-the-art performance, outperforming individual representations and competitive baselines.

\end{abstract}

\begin{comment}
In this work, we introduce the task of Synthetic Heart Sound Detection (SHAC), which targets the identification of phonocardiograms (PCGs) synthesized via neural audio codecs (NACs). PCGs or heart sounds act as a compelling biometric modality owing to their uniqueness, inherent liveness, and resilience against conventional spoofing attacks. However, rapid advances in NAC-based neural audio synthesis pose a serious threat: adversaries can now generate synthetic PCGs that are perceptually indistinguishable from genuine recordings and undermining the security of biometric authentication. To address this, we release the first benchmark dataset for SHAC, \textbf{\texttt{CARDIOFAKE}}, comprising both real and codec-synthesized PCGs, and conduct a comprehensive evaluation of spectral representations (MFCC, LFCC) and self-supervised learning (SSL) based pre-trained representations (e.g., WavLM). Building on these insights, we propose \textbf{\texttt{GROOT}}, a novel fusion framework that integrates SSL and spectral features for enhanced SHAC. By combining MFCC and WavLM, \textbf{\texttt{GROOT}} achieves state-of-the-art performance, surpassing individual representations and outperforming competitive baseline fusion methods.
\end{comment}

\section{Introduction \& Background}

Spoofing attack detection (SAD) is widely regarded as a core safeguard for biometric systems, and has been systematically explored in speech \cite{sanchez2015toward} and facial recognition \cite{li2018face}. In the speech domain, extensive research has investigated replay and voice conversion-based attacks, leading to standardized evaluations such as the ASVspoof challenges \cite{wu2015asvspoof,todisco2019asvspoof}. Through standardized protocols, these benchmarks have supported the development of effective countermeasures \cite{lavrentyeva2017audio,delgado2021asvspoof}. Work on face recognition has followed a similar path, from early texture-based methods to modern deep learning approaches \cite{maatta2011face, wen2015face, boulkenafet2015face}. Parallel advances have been made also in fingerprint and iris recognition \cite{menotti2015deep}. Collectively, these efforts underscore how community-driven efforts have catalyzed rapid progress in spoofing detection across established biometric systems. Despite continued progress in spoofing attack detection, the threat remains substantial. Conventional biometric modalities—including facial images, fingerprints, speech-have consistently been shown to be vulnerable to ever-growing advancement in sophisticated spoofing techniques. In contrast, phonocardiograms (PCGs), or heart sounds, have been regarded as a promising biometric modality, offering uniqueness, inherent liveness, and a natural resilience against traditional spoofing methods \cite{phua2008heart}. Unlike fingerprints, faces, or voices, heart sounds are directly tied to physiological processes, making them intrinsically difficult to forge. As such, subsequent research proposed a range of strategies for building biometric systems based on heart sounds. They have used wavelet-based features \cite{6577115}, cepstral features \cite{verma2014analysis} with classical ML algorithms to modern day transformer-based architectures \cite{10604926}. 

However, the very perception of heart sounds as a non-invasive biometric modality renders their vulnerability particularly concerning. With rapid advances in neural audio codec (NAC)–based synthesis, adversaries can now generate synthetic heart sounds that are perceptually indistinguishable from genuine recordings, posing a direct threat similar to those observed in synthetic speech generation \cite{lu24f_interspeech, wu24p_interspeech}. To confront this risk, we present the first systematic investigation into the vulnerability of heart sound biometrics under NAC-driven spoofing attacks—a new research direction for secure biometric authentication. As part of this effort for spoofing detection of such attacks- we coin the novel task of Synthetic Heart Sound Detection (SHAC) and release the first benchmark dataset, \textbf{\texttt{CARDIOFAKE}} (\textbf{\texttt{FAKE}} Phono\textbf{\texttt{CARDIO}}grams) comprising both real and codec-synthesized heart sounds. We conduct a comprehensive evaluation of spectral features (MFCC, LFCC) and self-supervised learning (SSL)–based representations (e.g., WavLM). \textit{We hypothesize that these two classes of features are complementary: spectral features are highly sensitive to NAC-induced distortions at the acoustic level, while SSL representations capture broader temporal structure and variability in heart sounds}. Building on this hypothesis, we introduce \texttt{\textbf{GROOT}} (Fusion via \texttt{\textbf{GR}}ammian \texttt{\textbf{O}}ptimal Transp\texttt{\textbf{O}}r\texttt{\textbf{T}}), a novel fusion framework that integrates spectral and SSL representations through novel grammian optimal transport. By combining MFCC with WavLM, \texttt{\textbf{GROOT}} achieves state-of-the-art (SOTA) performance, surpassing both individual representations and competitive baselines. Our benchmark and baselines establish a foundation for future research on robust countermeasures against this emerging class of attacks.

\noindent \textbf{Key contributions of this work are threefold:}  
\begin{itemize}
    \item We coin the novel task of SHAC as a new research direction for secure biometric authentication.
    \item We release the first benchmark dataset for SHAC, \textbf{\texttt{CARDIOFAKE}}, containing both real and codec-synthesized heart sounds, and conduct a comprehensive evaluation of spectral and SSL-based representations, providing critical insights into their strengths and limitations.
    \item We propose \texttt{\textbf{GROOT}}, a novel framework that integrates spectral and SSL representations to exploit their complementary strengths. At its core, \texttt{\textbf{GROOT}} employs a novel grammian optimal transport mechanism. \texttt{\textbf{GROOT}} achieves SOTA performance, outperforming individual representations and competitive baselines, thereby setting a strong foundation for future countermeasures against this emerging class of spoofing attacks. 
\end{itemize}
\noindent \textit{The dataset and codes are released here\footnote{\url{https://helixometry.github.io/SHAC/}}.}

\section{CARDIOFAKE Dataset}

This section outlines the resources and methodology employed in creating the \textbf{\texttt{CARDIOFAKE}} dataset, including the heart sounds corpora, the NACs used for synthesis, and the overall pipeline for producing the artificial samples. 

\subsection{Heart Sound Dataset}

For synthesizing heart sounds, we employ CirCor DigiScope dataset \cite{oliveira2021circor}, which is openly accessible via PhysioNet \cite{goldberger2000physiobank}. Our work concentrates on the open-access portion of the dataset, comprising recordings from 963 patients. Each patient record is labeled with one of three categories: Present, Absent, or Unknown. Altogether, the collection provides 3,163 phonocardiogram recordings, with durations spanning 5 to 65 seconds. 

\subsection{Neural Audio Codecs}

We utilize the NACs used by Lu et al. \cite{lu24f_interspeech} and Wu et al. \cite{wu24p_interspeech}, focusing on SOTA and publicly available codecs that adversaries might use for generation of synthetic heart sounds. The NACs leveraged are given as follows: \par

\noindent \textbf{Descript Audio Codec (DAC) \cite{kumar2024high}}: It is a high-fidelity VQ-GAN–based model that leverages residual vector quantization (RVQ) along with adversarial and multi-scale spectral losses; we employ its 16kHz variant. \par

\noindent \textbf{Encodec\cite{defossez2022high}}: It is a real-time convolutional encoder–decoder codec with RVQ, integrates time/frequency reconstruction losses and spectrogram adversarial objectives; we use the 24kHz version. \par

\noindent \textbf{Soundstream \cite{zeghidour2021soundstream}}: It is a NAC tailored for low-bitrate speech compression, employing an encoder–decoder framework with RVQ and multi-scale STFT discriminators to balance fidelity and compression efficiency, supporting bitrates from 3–18 kbps. We adopt its 16kHz variant. \par

\noindent \textbf{Speech Tokenizer \cite{zhang2024speechtokenizer}}: It serves as a unified audio tokenizer bridging semantic and acoustic cues. Built on an encoder–decoder with RVQ, it hierarchically disentangles content and paralinguistic information across layers, producing composite token sequences. We employ its default 16kHz configuration. \par

\noindent \textbf{FunCodec \cite{du2024funcodec}}: It incorporates RVQ with semantic augmentation and adversarial training to yield compact, expressive representations; we use the openly released LibriTTS trained and bilingual 16kHz version. \par

\noindent \textbf{AudioDec \cite{wu2023audiodec}}: It adopts an autoencoder framework with a two-stage training strategy—metric losses for convergence followed by decoder-only adversarial fine-tuning for fidelity; we employ the 28kHz variant. \par

\noindent \textbf{SNAC \cite{siuzdak2024snac}}: SNAC extends RVQ with hierarchical quantizers across temporal scales, depthwise convolutions, noise injection, and local attention; we use 24kHz version in our study.

\subsection{CARDIOFAKE Generation Pipeline}

We design a controlled pipeline inspired by prior work on NAC synthesized deepfakes \cite{wu24p_interspeech} for building \textbf{\texttt{CARDIOFAKE}}. We start with CirCor DigiScope dataset, where each utterance serves as a real reference. Synthetic samples are created through a NAC synthesis–resynthesis loop: the original waveform is first encoded into a discrete latent representation by a pre-trained NAC encoder and then reconstructed by its decoder, yielding a synthetic version. This process preserves the underlying cardiac acoustic patterns while introducing subtle codec-induced artifacts, resulting in realistic yet synthetic heart sounds. We apply this approach across 7 NACs, producing parallel datasets in which each real utterance has a one-to-one synthetic counterpart per codec. So, in total, we have 3163 real heart sounds and 22141 synthetic heart sounds. Two evaluation settings are defined: seen, where test utterances are generated using the same NACs as in training (SNAC, DAC, EnCodec, Soundstream, Speech Tokenizer), and unseen, where test utterances are synthesized with different codecs (FunCodec, AudioDec) to evaluate generalization.

\section{Methodology}

\subsection{Feature Extraction}

We employ MFCC\footnote{\url{https://librosa.org/doc/main/generated/librosa.feature.mfcc.html}} and and linear-frequency cepstral coefficients (LFCC\footnote{\url{https://spafe.readthedocs.io/en/latest/features/lfcc.html}}) as spectral representations. We extract 14-dimensional LFCC and 40-dimensional MFCC after average pooling. We employ different SOTA SSL representations as they have shown effectiveness for heart sound classfication tasks \cite{panah2023exploring}. We use Wav2vec2\footnote{\url{https://huggingface.co/facebook/wav2vec2-base}} \cite{baevski2020wav2vec} pre-trained on 960 hours of Librispeech. Its self-supervised framework learns contextualized representations by masking speech frames and predicting quantized latent targets. Lastly, we use SOTA models in SUPERB i.e. Unispeech-SAT\footnote{\url{https://huggingface.co/microsoft/unispeech-sat-base}} \cite{chen2022unispeech} and WavLM\footnote{\url{https://huggingface.co/microsoft/wavlm-base}} \cite{chen2022wavlm}. Unispeech-SAT extends self-supervised learning by incorporating speaker-aware objectives.This enables the model to disentangle speaker identity from linguistic content, whereas WavLM incorporates masked prediction with denoising objectives. We resample all the inputs to 16 kHz and extract representations by average pooling over the final hidden layer of each frozen SSL model. We extract 768 dimension representation for all the SSL models.

\subsection{Individual Representations Modeling}

We follow previous work and experiment with two downstream modeling architectures that shown SOTA results on audio deepfake detection with SSL representations \cite{chetia-phukan-etal-2024-heterogeneity}. Firstly, we apply a fully connected network (FCN) with two dense layers of 180 and 60 neurons to the extracted representations. Secondly, we use CNN, we place a 1D-CNN layer with 32 filters on top of the representations, followed by a max-pooling layer, flatten and a FCN with the same details as the FCN above. The output layer of both models uses a sigmoid activation function. 

\begin{figure}[!hbt]
    \centering
    \includegraphics[scale=0.15]{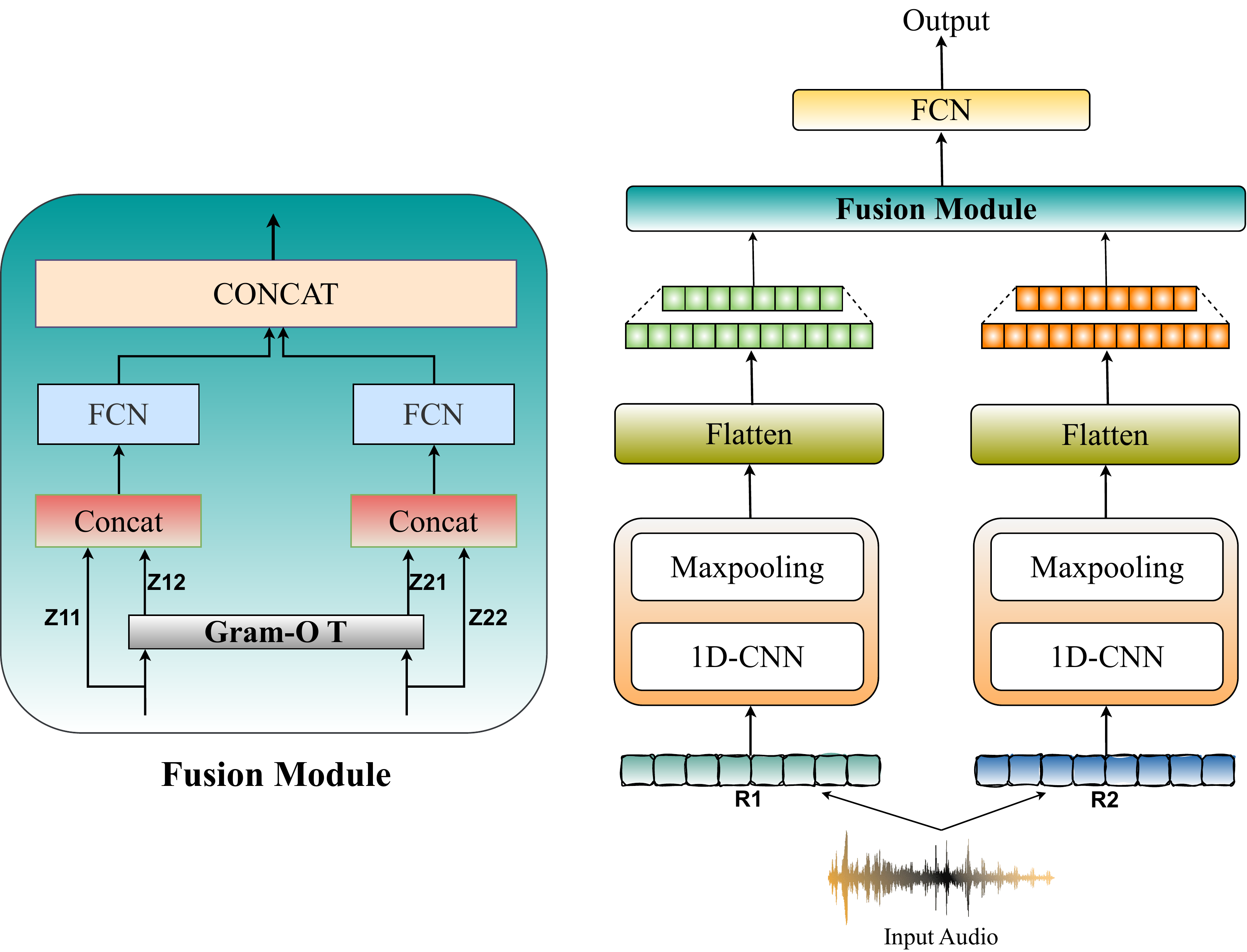}
    \caption{Proposed Framework: \textbf{\texttt{GROOT}}: R1 and R2 represent input features from two branches. $Z_{11}$ and $Z_{22}$ denote features from the respective FCN branches, while $Z_{12}$ and $Z_{21}$ denote transported features}
    \label{fig:work_flow}
\end{figure}

\subsection{GROOT}

We propose \texttt{\textbf{GROOT}} for the fusion of representations and the architecture is shown in Figure~\ref{fig:work_flow}. The extracted representations are first passed through a 1D-CNN block with the same design as used in the individual CNN models. The output is then flattened and linearly projected to a 120-dimensional vector, where the dimensionality reduction is introduced mainly to reduce computational cost. These projected features are then fed into the fusion module, which employs the proposed novel grammian optimal transport (Gram-OT) for aligning the representations. Vanilla optimal transport (OT) has been widely adopted for representation alignment \cite{pramanick2022multimodal, rho2025lavcap}, but it directly compares raw features, making it sensitive to scaling and noisy variations. In contrast, Gram-OT compares representations through their gram matrices, which capture correlations between features and reflect global relational patterns across the representation space. This enables Gram-OT to preserve meaningful characteristics such as rhythm while being more robust to noise, distortions, and variability across inputs. Let us consider the feature vectors of two representations after flattening are $R_1$ and $R_2$. We first compute their gram matrices: 
\[
 G_{R_1} = R_1 R_1^\top \quad G_{R_2} = R_2 R_2^\top 
\]
\noindent We then construct a cost matrix based on the frobenius distance between the two gram matrices:
\[
M
= \frac{\left\lVert G_{R_1}-G_{R_2}\right\rVert_F}
{\displaystyle \max_{(R_1,R_2)} \left\lVert G_{R_1}-G_{R_2}\right\rVert_F }
\]

\noindent Here, frobenius distance is used in place of euclidean distance used in vanilla OT, as it provides the natural generalization of euclidean distance from vectors to matrices, making it well-suited for comparing gram matrices. To align the features, we apply the sinkhorn algorithm on this cost to obtain the optimal transport plan $\Gamma$
\[
\Gamma = \text{Sinkhorn}(M) 
\]
\noindent Using $\Gamma$, we transport the feature spaces of the two representations into one another: 

\[
R_2 \rightarrow R_1 = \Gamma \cdot R_2 \quad R _1 \rightarrow R_2 = \Gamma^\top \cdot R_1
\]

\noindent Finally, the transported features are concatenated with their corresponding original representations to form the fused representations $F_1$ and $F_2$: 
\[
F_1 = \text{Concat}(R_2 \rightarrow R_1, R_1) \quad F_2 = \text{Concat}(R_1 \rightarrow R_2, R_2)
\]
\noindent The fused representations $F_1$ and $F_2$ are first passed in parallel through FCN with a dense layer of 80 neurons each, and their outputs are subsequently concatenated. This concatenated vector is then processed by another FCN consisting of two dense layers with 120 and 30 neurons, respectively, followed by the final output layer with a sigmoid activation function for binary classification.

\section{Experiments}

\subsection{Training and Hyperparameter Details}

All models are trained for 50 epochs with a batch size of 32, using the Adam optimizer and binary cross-entropy loss. To mitigate overfitting, dropout regularization is applied. We keep a learning rate of 1e-3 for the experiments. We also use class-weightage during training to handle the class-imbalance. 

\subsection{Experimental Results}

We begin by evaluating whether NAC-synthesized heart sounds retain patient-specific identity, thereby assessing the credibility of such spoofing attacks. To this end, we perform a closed-set user identification experiment on subjects from the real heart sound corpus. Each patient is treated as a distinct class, and a supervised classifier is trained under four train–test regimes: Real→Real, Real→Fake, Fake→Real, and Fake→Fake (where Real = authentic heart sound, Fake = NAC-synthesized heart sound). The classifier achieves ~89.11\% accuracy in the Real→Real setting, confirming that genuine heart sounds reliably encode identity cues. Importantly, the Real→Fake regime still reaches 86.29\%, indicating that synthetic reconstructions preserve most of the patient-specific information. Moreover, models trained on synthetic data perform even better: 95.07\% in Fake→Fake and 93.08\% in Fake→Real. These results demonstrate that NAC-based synthesis preserves discriminative identity cues and the higher accuracy is due to the more samples in the synthetic set. In summary, NAC-generated heart sounds constitute highly identity-preserving deepfakes, posing a critical challenge for biometric systems that may struggle to differentiate authentic from synthetic.

\begin{table}[!hbt]
\centering
\setlength{\tabcolsep}{8pt}
\begin{tabular}{l|cc|cc}
\toprule
\multirow{2}{*}{\textbf{PTM's}} & \multicolumn{2}{c|}{\textbf{FCN}} & \multicolumn{2}{c}{\textbf{CNN}} \\
\cmidrule(lr){2-3} \cmidrule(lr){4-5}
 & \textbf{ACC \(\uparrow\)} & \textbf{EER \(\downarrow\)} & \textbf{ACC \(\uparrow\)} & \textbf{EER \(\downarrow\)} \\
\midrule
\multicolumn{5}{c}{\textbf{Seen}} \\
\midrule
\textbf{LF}     & \cellcolor{lavender!45}76.99 & \cellcolor{lavender!45}15.19 & \cellcolor{lavender!65}79.02 & \cellcolor{lavender!45}14.96 \\
\textbf{MF}     & \cellcolor{lavender!45}77.82 & \cellcolor{lavender!45}15.04 & \cellcolor{lavender!65}81.56 & \cellcolor{lavender!80}12.55 \\
\textbf{W2V}    & \cellcolor{lavender!65}83.62 & \cellcolor{lavender!65}\textbf{12.13} & \cellcolor{lavender!80}86.65 & \cellcolor{lavender!80}10.37 \\
\textbf{UNS}    & \cellcolor{lavender!65}80.30 & \cellcolor{lavender!65}12.30 & \cellcolor{lavender!65}82.81 & \cellcolor{lavender!80}11.59 \\
\textbf{WAL}    & \cellcolor{lavender!80}\textbf{84.54} & \cellcolor{lavender!65}12.51 & \cellcolor{lavender!80}\textbf{87.72} & \cellcolor{lavender!90}\textbf{9.45} \\
\midrule
\multicolumn{5}{c}{\textbf{Unseen}} \\
\midrule
\textbf{LF}     & \cellcolor{lavender!45}72.45 & \cellcolor{lavender!25}18.93 & \cellcolor{lavender!45}73.99 & \cellcolor{lavender!25}18.08 \\
\textbf{MF}     & \cellcolor{lavender!45}74.93 & \cellcolor{lavender!30}17.60 & \cellcolor{lavender!45}78.74 & \cellcolor{lavender!30}16.91 \\
\textbf{W2V}    & \cellcolor{lavender!65}79.56 & \cellcolor{lavender!30}16.03 & \cellcolor{lavender!65}83.61 & \cellcolor{lavender!65}13.74 \\
\textbf{UNS}    & \cellcolor{lavender!45}74.02 & \cellcolor{lavender!25}18.07 & \cellcolor{lavender!45}78.47 & \cellcolor{lavender!25}18.69 \\
\textbf{WAL}    & \cellcolor{lavender!65}\textbf{80.54} & \cellcolor{lavender!45}\textbf{15.01} & \cellcolor{lavender!80}\textbf{84.02} & \cellcolor{lavender!65}\textbf{13.39} \\
\bottomrule
\end{tabular}
\caption{Accuracy (ACC) and Equal Error Rate (EER) for Seen and Unseen conditions; Abbreviation used: LFCC (LF), MFCC (MF), Wav2vec2 (W2V), Unispeech-SAT (UNS), WavLM (WAL). All scores are in \%. \textit{Abbreviations are consistent in Table~\ref{2}.}}
\label{baseline}
\end{table}

\noindent Table~\ref{baseline} presents the results of spectral and SSL features with FCN and CNN downstream models under both seen and unseen evaluation conditions. Overall, CNN-based downstream models consistently outperform their FCN counterparts. Among individual representations, WavLM with CNN emerges as the strongest performer, surpassing Wav2vec2 and Unispeech-SAT across both conditions. Furthermore, SSL features consistently outperform spectral counterparts (MFCC and LFCC), highlighting their effectiveness. Table~\ref{2} reports the results of representation fusion. We compare against two baselines: simple concatenation and optimal transport (OT) \cite{pramanick2022multimodal}. For a fair comparison, we retain the same architecture as \textbf{\texttt{GROOT}} for OT, differing only in the computation of the gram matrix. Similarly, all training settings are kept identical to those used in \textbf{\texttt{GROOT}} for both concatenation and OT.We observe that fusion of representations through \textbf{\texttt{GROOT}} consistently achieves the best overall performance. Moreover, a clear trend emerges: heterogeneous fusion of spectral and SSL representations outperforms homogeneous fusion, thereby validating our hypothesis on the complementarity of these feature classes. We observe the best performance with fusion of MFCC and WavLM through \textbf{\texttt{GROOT}}. These findings not only establish strong baselines for this novel task but also open new avenues for future research.

\begin{table}[!hbt]
\centering
\setlength{\tabcolsep}{2pt}
\begin{tabular}{l|cc|cc|cc}
\toprule
\multirow{2}{*}{\textbf{Fusion}} & \multicolumn{2}{c|}{\textbf{Concat}} & \multicolumn{2}{c|}{\textbf{OT}} & \multicolumn{2}{c}{\textbf{\texttt{GROOT}}} \\
\cmidrule(lr){2-3} \cmidrule(lr){4-5} \cmidrule(lr){6-7}
 & \textbf{ACC \(\uparrow\)} & \textbf{EER \(\downarrow\)} & \textbf{ACC \(\uparrow\)} & \textbf{EER \(\downarrow\)} & \textbf{ACC \(\uparrow\)} & \textbf{EER \(\downarrow\)} \\
\midrule
\multicolumn{7}{c}{\textbf{Seen}} \\
\midrule
\textbf{LF + MF}    & \cellcolor{lavender!30}80.05 & \cellcolor{lavender!35}11.82 & \cellcolor{lavender!45}82.41 & \cellcolor{lavender!50}10.91 & \cellcolor{lavender!65}84.61 & \cellcolor{lavender!70}8.35 \\
\textbf{LF + W2V}   & \cellcolor{lavender!40}86.18 & \cellcolor{lavender!50}8.72  & \cellcolor{lavender!55}88.93 & \cellcolor{lavender!60}8.10  & \cellcolor{lavender!75}90.50 & \cellcolor{lavender!80}7.42 \\
\textbf{LF + UNS}   & \cellcolor{lavender!35}81.12 & \cellcolor{lavender!40}12.00 & \cellcolor{lavender!50}84.50 & \cellcolor{lavender!45}11.18 & \cellcolor{lavender!65}87.09 & \cellcolor{lavender!60}9.63 \\
\textbf{LF + WAL}   & \cellcolor{lavender!40}86.32 & \cellcolor{lavender!55}7.18  & \cellcolor{lavender!55}88.36 & \cellcolor{lavender!60}7.03  & \cellcolor{lavender!75}91.77 & \cellcolor{lavender!80}6.06 \\
\textbf{MF + W2V}   & \cellcolor{lavender!40}86.57 & \cellcolor{lavender!55}7.20  & \cellcolor{lavender!55}88.82 & \cellcolor{lavender!60}6.87  & \cellcolor{lavender!70}90.83 & \cellcolor{lavender!75}6.14 \\
\textbf{MF + UNS}   & \cellcolor{lavender!35}84.00 & \cellcolor{lavender!40}11.23 & \cellcolor{lavender!50}85.88 & \cellcolor{lavender!45}11.22 & \cellcolor{lavender!65}87.90 & \cellcolor{lavender!60}9.72 \\
\textbf{MF + WAL}   & \cellcolor{lavender!40}\textbf{87.70} & \cellcolor{lavender!55}\textbf{7.40}  & \cellcolor{lavender!55}\textbf{89.07} & \cellcolor{lavender!60}\textbf{6.86}  & \cellcolor{lavender!75}\textbf{93.20} & \cellcolor{lavender!80}\textbf{5.86} \\
\textbf{W2V + UNS}  & \cellcolor{lavender!40}86.99 & \cellcolor{lavender!45}9.87  & \cellcolor{lavender!55}87.79 & \cellcolor{lavender!50}9.06  & \cellcolor{lavender!65}89.00 & \cellcolor{lavender!55}8.61 \\
\textbf{W2V + WAL}  & \cellcolor{lavender!40}86.26 & \cellcolor{lavender!50}8.32  & \cellcolor{lavender!55}88.92 & \cellcolor{lavender!55}7.82  & \cellcolor{lavender!70}91.60 & \cellcolor{lavender!75}6.20 \\
\textbf{UNS + WAL}  & \cellcolor{lavender!35}85.01 & \cellcolor{lavender!40}10.54 & \cellcolor{lavender!50}86.23 & \cellcolor{lavender!45}8.92  & \cellcolor{lavender!65}88.74 & \cellcolor{lavender!55}7.55 \\
\midrule
\multicolumn{7}{c}{\textbf{Unseen}} \\
\midrule
\textbf{LF + MF}    & \cellcolor{lavender!25}79.28 & \cellcolor{lavender!20}16.70 & \cellcolor{lavender!35}81.87 & \cellcolor{lavender!25}15.22 & \cellcolor{lavender!50}83.70 & \cellcolor{lavender!35}13.99 \\
\textbf{LF + W2V}   & \cellcolor{lavender!35}84.11 & \cellcolor{lavender!30}12.34 & \cellcolor{lavender!50}84.25 & \cellcolor{lavender!40}10.83 & \cellcolor{lavender!60}86.00 & \cellcolor{lavender!45}9.87 \\
\textbf{LF + UNS}   & \cellcolor{lavender!25}79.05 & \cellcolor{lavender!15}17.98 & \cellcolor{lavender!35}80.47 & \cellcolor{lavender!20}16.00 & \cellcolor{lavender!50}82.39 & \cellcolor{lavender!30}14.80 \\
\textbf{LF + WAL}   & \cellcolor{lavender!30}81.22 & \cellcolor{lavender!25}13.27 & \cellcolor{lavender!45}83.01 & \cellcolor{lavender!35}12.87 & \cellcolor{lavender!60}85.31 & \cellcolor{lavender!40}10.98 \\
\textbf{MF + W2V}   & \cellcolor{lavender!30}81.00 & \cellcolor{lavender!25}\textbf{12.08} & \cellcolor{lavender!45}84.72 & \cellcolor{lavender!35}\textbf{10.34} & \cellcolor{lavender!60}85.78 & \cellcolor{lavender!40}10.70 \\
\textbf{MF + UNS}   & \cellcolor{lavender!30}80.72 & \cellcolor{lavender!20}14.89 & \cellcolor{lavender!40}81.51 & \cellcolor{lavender!25}13.70 & \cellcolor{lavender!55}83.04 & \cellcolor{lavender!30}11.49 \\
\textbf{MF + WAL}   & \cellcolor{lavender!35}\textbf{84.33} & \cellcolor{lavender!25}13.11 & \cellcolor{lavender!50}\textbf{84.99} & \cellcolor{lavender!35}12.06 & \cellcolor{lavender!60}\textbf{86.10} & \cellcolor{lavender!40}\textbf{9.75} \\
\textbf{W2V + UNS}  & \cellcolor{lavender!35}83.11 & \cellcolor{lavender!25}13.38 & \cellcolor{lavender!45}84.20 & \cellcolor{lavender!30}12.40 & \cellcolor{lavender!60}85.81 & \cellcolor{lavender!35}11.13 \\
\textbf{W2V + WAL}  & \cellcolor{lavender!35}83.97 & \cellcolor{lavender!25}12.90 & \cellcolor{lavender!45}84.09 & \cellcolor{lavender!35}11.29 & \cellcolor{lavender!60}85.70 & \cellcolor{lavender!40}10.00 \\
\textbf{UNS + WAL}  & \cellcolor{lavender!30}81.02 & \cellcolor{lavender!20}15.80 & \cellcolor{lavender!40}82.68 & \cellcolor{lavender!25}14.10 & \cellcolor{lavender!55}84.48 & \cellcolor{lavender!30}12.51 \\
\bottomrule
\end{tabular}
\caption{Evaluation scores for fusion of features; All scores are in \%; OT stands for Optimal Transport}
\label{2}
\end{table}

\subsection{Comparison to SOTA}

\begin{figure}[!hbt]
    \centering
    \subfloat[MiO]{%
        \includegraphics[width=0.23\textwidth]{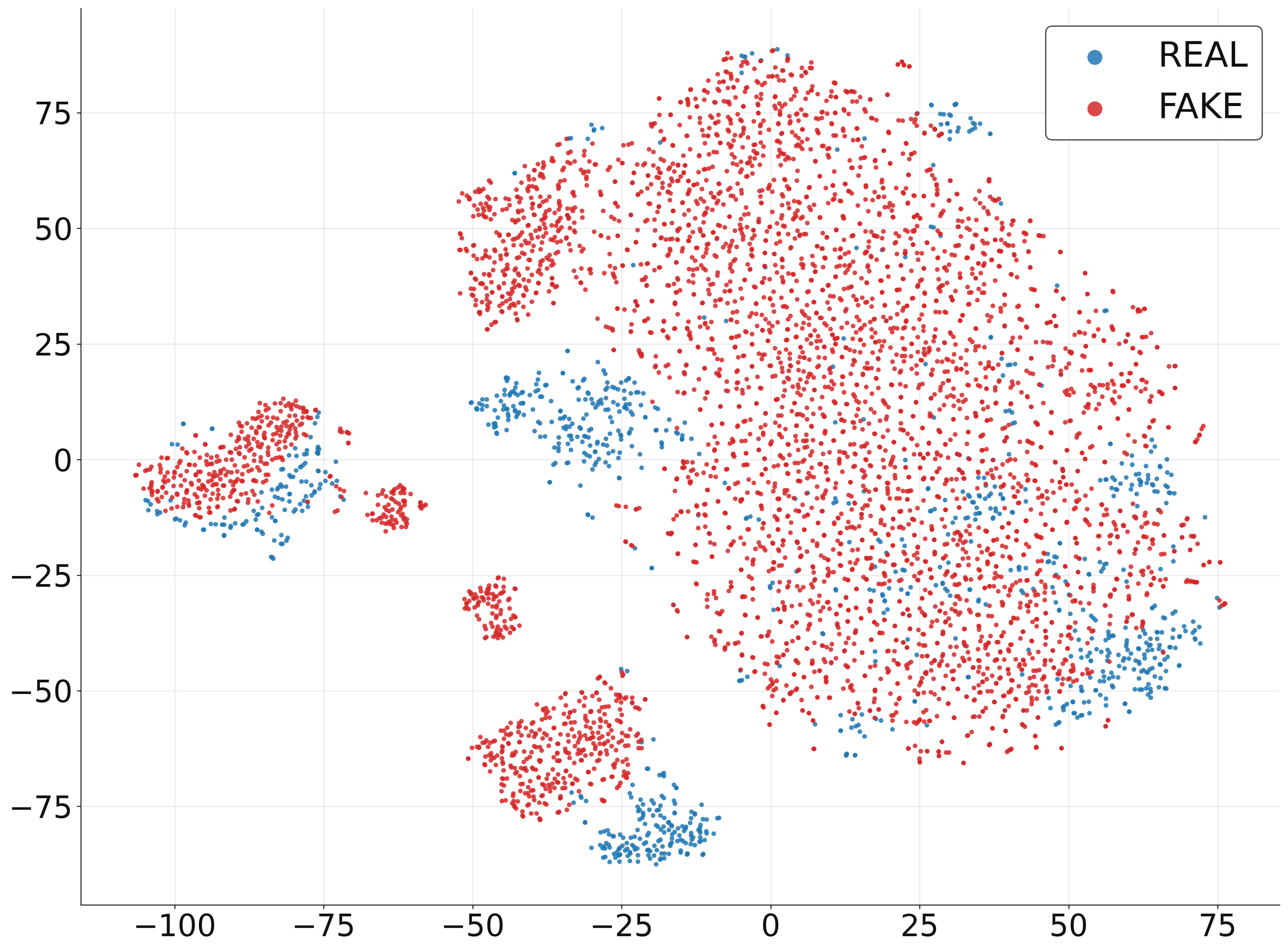}
    }
    \hfill
    \subfloat[\textbf{\texttt{GROOT (MFCC + WavLM)}}]{%
        \includegraphics[width=0.23\textwidth]{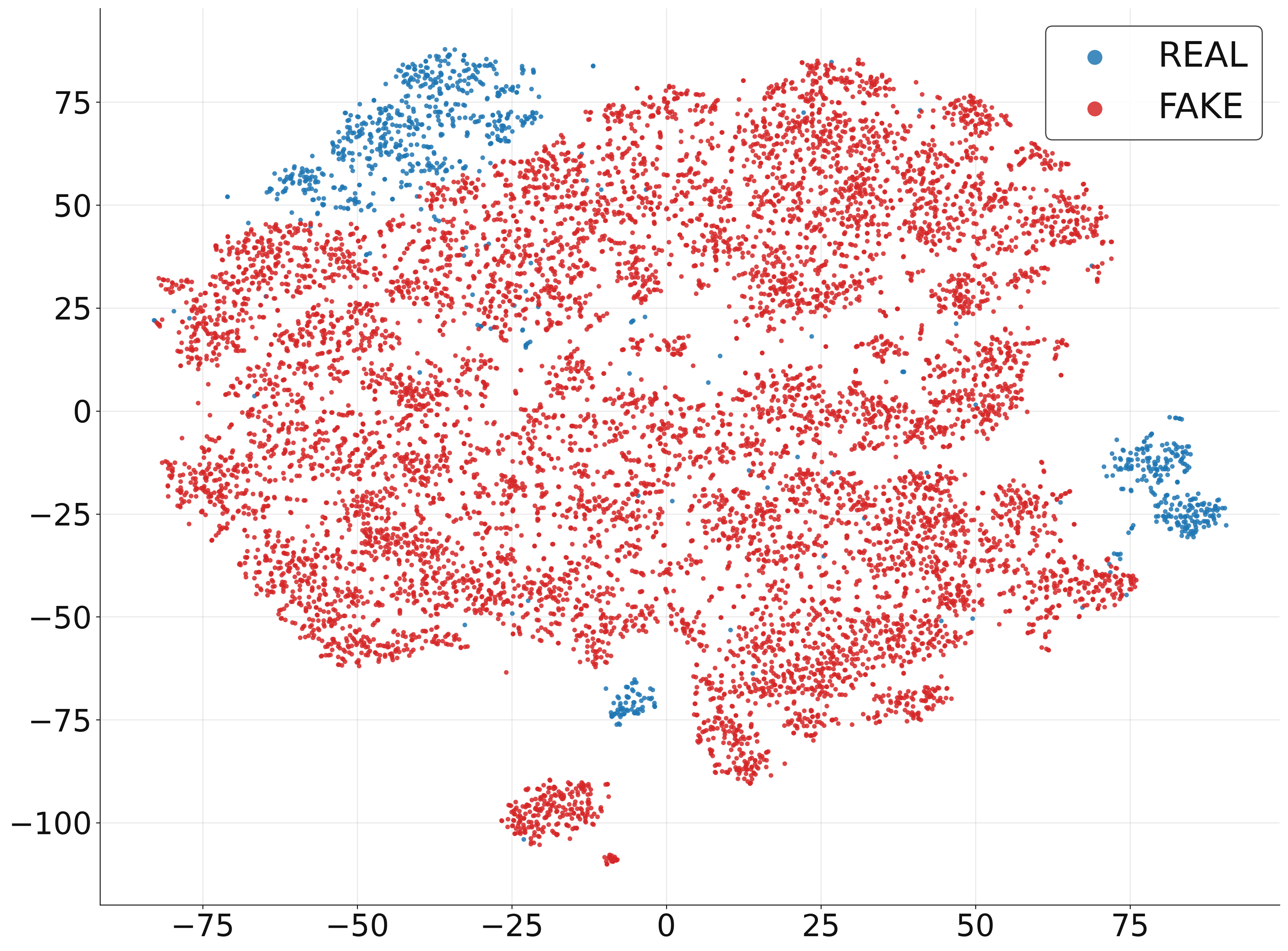}
    }

    \caption{t-SNE plots}
    \label{fig:tsne}
\end{figure}

\begin{figure}[!hbt]
    \centering
    \subfloat[MiO]{%
        \includegraphics[width=0.23\textwidth]{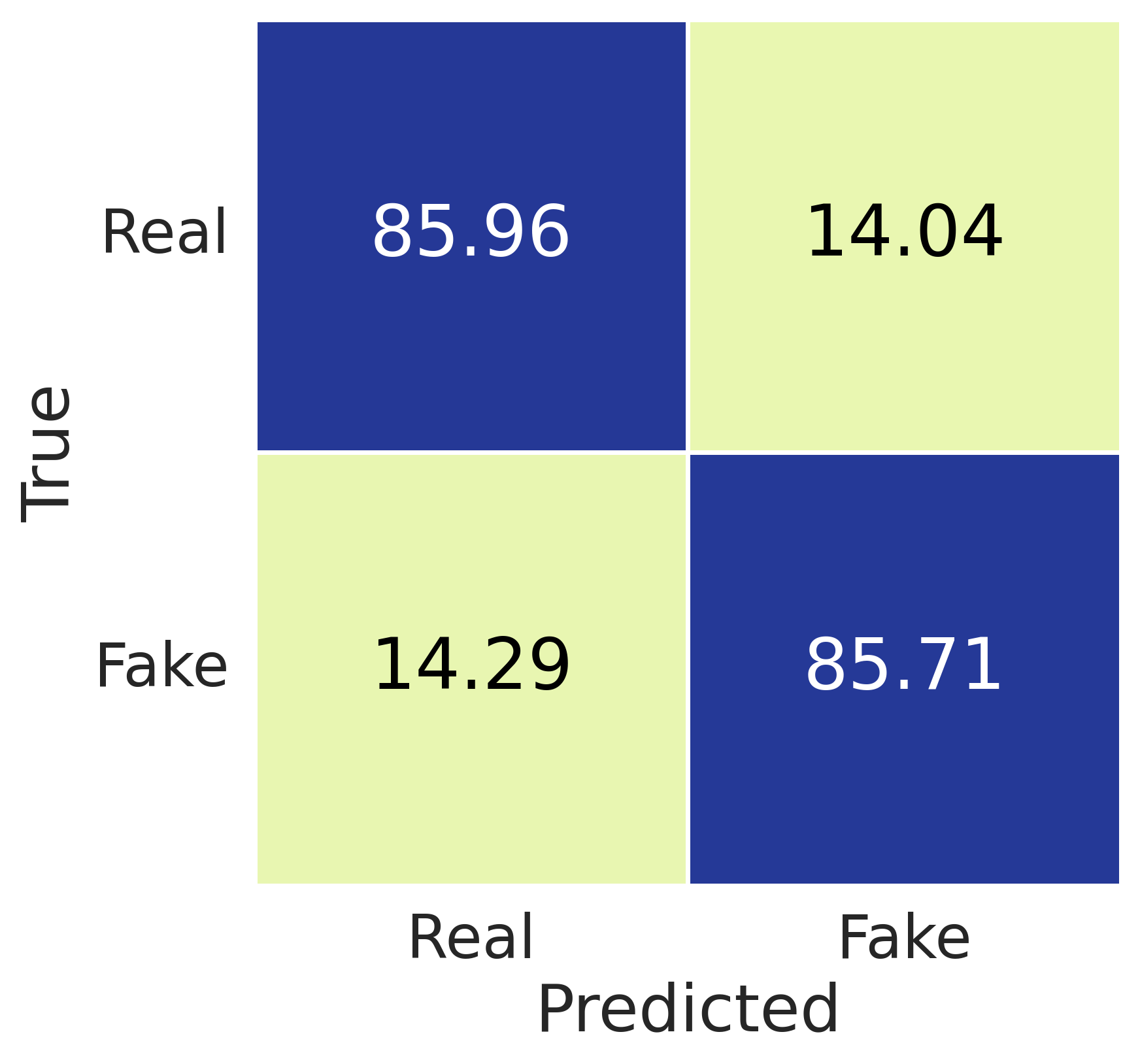}
    }
    \hfill
    \subfloat[\textbf{\texttt{GROOT (MFCC + WavLM)}}]{%
        \includegraphics[width=0.23\textwidth]{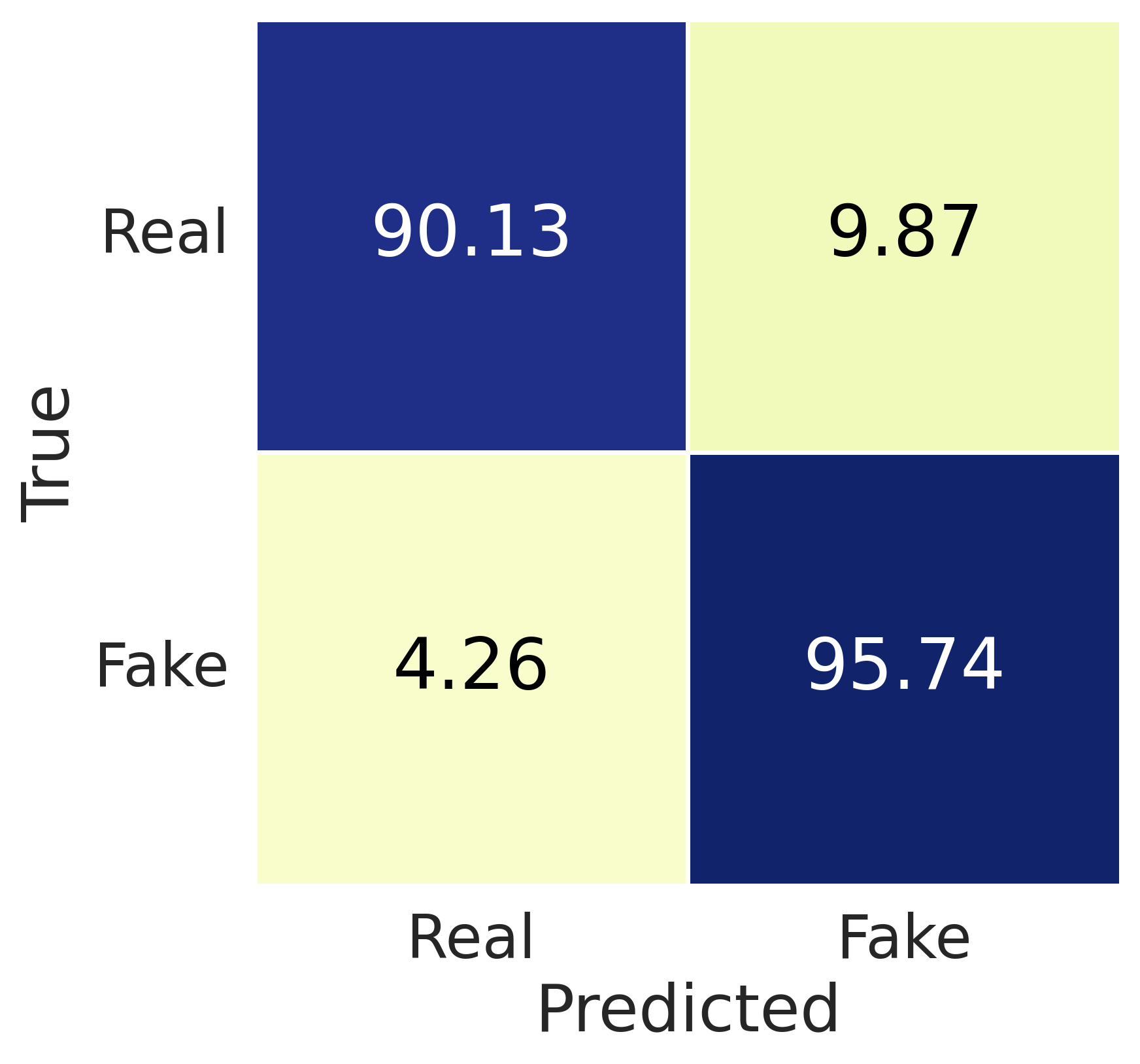}
    }

    \caption{Confusion Matrices}
    \label{fig:cm}
\end{figure}

As this is the first work addressing SHAC, there are currently no task-specific SOTA models for direct comparison. Therefore, we compare our best-performing method, \textbf{\texttt{GROOT}} with MFCC + WavLM (See Table \ref{2}), against strong baselines from general audio deepfake detection, namely AASIST \cite{jung2022aasist} and MiO \cite{phukan2024heterogeneity}. AASIST is a graph neural network-based architecture, while MiO performs outer-product fusion of SSL representations, making them competitive baselines for this task. We train them following the training configuration for \textbf{\texttt{GROOT}} for fair comparison. Our proposed \textbf{\texttt{GROOT}} achieves 93.20\% accuracy and 5.86\% EER in the Seen setting, and 86.10\% accuracy with 9.75\% EER in the Unseen setting. In comparison, AASIST achieves 85.15\% accuracy and 14.91\% EER (Seen) and 73.13\% accuracy with 16.43\% EER (Unseen). MiO shows slight improvement over AASIST with 86.98\% accuracy and 12.34\% EER (Seen) and 75.89\% accuracy with 14.09\% EER (Unseen).We further visualize the learned representations using t-SNE plots from the penultimate layers of MiO and \textbf{\texttt{GROOT}} (MFCC + WavLM) in Fig. \ref{fig:tsne}. The visualization shows clearer separation and tighter clustering between real and fake heart sounds for \textbf{\texttt{GROOT}}, indicating more discriminative representations. Additionally, the confusion matrices for both models are shown in Fig. \ref{fig:cm}, where \textbf{\texttt{GROOT}} demonstrates fewer misclassifications compared to MiO. These observations further confirm that \textbf{\texttt{GROOT}} significantly outperforms strong audio deepfake baselines, establishing it as a strong baseline for the SHAC task.

\section{Conclusion}  
In summary, this work introduces the novel task of SHAC, highlighting the emerging risk posed by NACs. To facilitate research in this direction, we release \textbf{\texttt{CARDIOFAKE}}, the first benchmark dataset for SHAC, containing both real and codec-synthesized heart sounds. We perform extensive evalaution of both spectral (MFCC, LFCC) and SSL representations (e.g., WavLM) for SHAC. Finally, we present \textbf{\texttt{GROOT}}, a novel framework that effectively integrates SSL and spectral features by exploiting their complementary behavior, setting a SOTA for SHAC by outperforming both individual representations and strong competitive baselines.

{
\section{Generative AI Use Disclosure}
AI-assisted tools were used only to enhance grammar, clarity, and overall presentation of the manuscript. These tools were not involved in developing the scientific ideas, conducting data analysis, generating results, or interpreting the findings. The authors take full responsibility for the accuracy, validity, and integrity of the work.}

\bibliographystyle{IEEEtran}
\bibliography{mybib}

\end{document}